\documentclass[aps,prb,twocolumn,groupedaddress,showpacs,floatfix]{revtex4-1}

\usepackage{graphicx}
\usepackage{dcolumn}
\usepackage{bm}
\usepackage{amsmath}
\DeclareMathOperator{\Tr}{Tr}

\begin{document}

\title{Interplay of relativistic and nonrelativistic transport in atomically
precise segmented graphene nanoribbons}

\author{Constantine Yannouleas}
\author{Igor Romanovsky}
\author{Uzi Landman}

\affiliation{School of Physics, Georgia Institute of Technology,
             Atlanta, Georgia 30332-0430}

\date{14 September 2014}

\begin{abstract}
Graphene's isolation launched explorations of fundamental relativistic physics originating from
the planar honeycomb lattice arrangement of the carbon atoms, and of potential
technological applications in nanoscale electronics. Bottom-up fabricated atomically-precise
segmented graphene nanoribbons, SGNRs, open avenues for studies of electrical transport, coherence,
and interference effects in metallic, semiconducting, and mixed GNRs, with different
edge terminations. Conceptual and practical understanding of electric transport through SGNRs
is gained through nonequilibrium Green's function (NEGF) conductance calculations and a
Dirac continuum model that absorbs the valence-to-conductance energy gaps as position-dependent
masses, including topological-in-origin mass-barriers at the contacts between segments. The continuum
model reproduces the NEGF results, including optical Dirac Fabry-P\'{e}rot (FP) equidistant
oscillations for massless relativistic carriers in metallic armchair SGNRs, and an unequally-spaced
FP pattern for mixed armchair-zigzag SGNRs where carriers transit from a relativistic (armchair)
to a nonrelativistic (zigzag) regime. This provides a unifying framework for analysis of coherent
transport phenomena and interpretation of forthcoming experiments in SGNRs.
\end{abstract}

\maketitle

Graphene, a single-atom-thin plane of graphite, has been the focus of intensive research endeavors 
since its isolation in 2004.\cite{geim04} The high degree of interest in this material originates from
its outstanding electronic, mechanical, and physical properties that result from the planar arrangement of the 
carbon atoms in a honeycomb lattice. Indeed graphene is considered both as a vehicle for exploring fundamental
relativistic physics, as well as a promising material for potential technological applications in
nanoscale electronics and optics.\cite{razabook} 

However, the absence of an electronic energy gap between the valence and conduction bands of 2D graphene casts 
doubts on its use in nanoelectronic devices. Nevertheless, theoretical studies had predicted that narrow graphene
nanoribbons (GNRs) can have a large band gap, comparable to silicon ($\sim$ 1 eV), depending on the ribbon's 
width $W$ and edge geometry (as well as possible doping at controlled positions). Pertinent to our work, we
note that these predictions were made \cite{fuji96,cohe06,scus06,wang07,waka10} for GNRs that have atomically 
precise armchair edges with widths $W \leq 2$ nm. Consequently, the most recent advent and growing availability 
of bottom-up fabricated atomically-precise narrow graphene nanoribbons, 
\cite{ruff10,huan12,derl13,nari14,sini14} including segmented \cite{ruff12} armchair graphene 
nanoribbons (SaGNRs), opens promising avenues for graphene nanoelectronics and for detailed explorations of 
coherent electrical transport in nanoribbon-based graphene wires, nanoconstrictions, and quantum-point contacts. 

Here we report on the unique apects of transport through segmented GNRs 
obtained from tight-binding non-equilibrium Green's function \cite{dattabook} (TB-NEGF) calculations in
conjunction with an analysis based on a one-dimensional (1D) relativistic Dirac model. 
This model is referred to by us as the Dirac-Fabry-P\'{e}rot (DFP) theory (see below for the choice of name). In
particular, it is shown that the valence-to-conduction energy gap in armchair GNR (aGNR) segments, as well as the
barriers at the interfaces between nanoribbon segments, can be incorporated in an effective position-dependent mass
term in the Dirac hamiltonian; the transport solutions associated with this hamiltonian exhibit conductance
patterns comparable to those obtained from the microscopic NEGF calculations. For zigzag graphene nanoribbon (zGNR)
segments, the valence-to-conduction energy gap vanishes, and the mass term is consonant with excitations
corresponding to massive nonrelativistic Schr\"{o}dinger-type carriers.

As aforementioned,
transport through narrow graphene channels $-$ particularly bottom-up fabricated and atomically-precise 
graphene nanoribbons \cite{ruff10,huan12,derl13,nari14,sini14,ruff12} $-$ is expected to offer ingress 
to unique behavior of Dirac electrons in graphene nanostructures. In particular,
the wave nature of elementary particles (e.g., electrons and photons) is commonly manifested and demonstrated
in transport processes. Because of an exceptionally high electron mobility and a long mean-free path, \cite{geim04} 
it has been anticipated that graphene devices hold the promise for the realization, measurement, and possible
utilization of fundamental aspects of coherent and ballistic transport behavior, which to date have been
observed, with varying degrees of success, mainly at semiconductor interfaces,\cite{hou89,ji03} quantum point 
contacts,\cite{vwee88} metallic wires,\cite{pasc95} and carbon nanotubes.\cite{lian01} Prominent among 
the effects that accompany coherence and ballistic transport are conductance quantization (in 
nanoconstrictions) in steps of $G_0=2e^2/h$, which have been found earlier
for quantum ballistic transport in semiconductor point contacts\cite{vwee88} and metal nanowires.\cite{pasc95} 
However, quantization signatures were scarcely observed \cite{vwee11} in GNRs fabricated with top-down methods.

Another manifestation of coherent ballistic transport are interference phenomena, reflecting the 
wave nature of the transporting physical object, and associated most often with optical (electromagnetic 
waves, photons) systems or with analogies to such systems (that is, the behavior of massless particles, as 
in graphene sheets). Measurements of interference patterns are commonly made with the use of interferometers, 
most familiar among them the multi-pass optical Fabry-P\'{e}rot (OFP) interferometer,\cite{lipsbook}
where the superposition of all the outgoing light waves, bouncing in a cavity from bounding partially 
reflecting mirrors, yields an oscillatory intensity record (interference pattern) which depends on
the light wavelength and the distance between the mirrors.

The quest for demonstration of the particle-wave duality of electrons through measurement of quantum
interference phenomena associated with electron transport in solid-state devices requires materials and
configurations having  long mean-free paths. This requirement limited early experimental work in the late 
1980s to semiconductor heterostructures,\cite{hou89,ji03} where conductance-quantization steps in 
gate-controlled two-dimensional constrictions have been observed. Subsequently interference patterns for 
nonrelativistic charge carriers in the form of conductance oscillations were observed\cite{lieb05} (and 
interpreted \cite{lieb05,land00} as Fabry-P\'{e}rot-type phenomena) in semiconductor nanowires.
The advent of 2D forms of carbon allotropes has motivated the study of optical-like interference
phenomena associated with relativistic massless electrons, as in the case of metallic carbon nanotubes
\cite{lian01} and graphene 2D $p$-$n$ junctions.\cite{rick13} (We note that the hallmark of the OFP is that 
the energy separation between successive maxima of the interference pattern varies as the inverse of the cavity
length $L$.)

For GNRs with segments of different widths, our investigations reveal diverse transport modes beyond the
OFP case, with conductance quantization steps ($nG_0, n = 1, 2, 3, \ldots$) found only for uniform GNRs. 
In particular, three distinct categories of Fabry-P\'{e}rot interference patterns are identified:
\begin{enumerate}
\item
{\it FP-A:\/} An {\it optical\/} FP pattern corresponding to {\it massless\/} graphene electrons 
exhibiting {\it equal spacing\/} between neighboring peaks. This pattern is associated with 
metallic armchair nanoribbon central segments. This category is subdivided further to {\it FP-A1\/} and 
{\it FP-A2} depending on whether a valence-to-conduction gap is absent ({\it FP-A1}, associated with 
metallic armchair leads), or present ({\it FP-A2}, corresponding to semiconducting armchair leads).
\item
{\it FP-B:} A {\it massive relativistic\/} FP pattern exhibiting a shift in the conduction onset due to the
valence-to-conduction gap and {\it unequal peak spacings\/}. This pattern is associated with semiconducting
armchair nanoribbon central segments, irrespective of whether the leads are metallic armchair, semiconducting
armchair, or zigzag.
\item
{\it FP-C:} A {\it massive non-relativistic\/} FP pattern with $1/L^2$ {\it peak spacings\/}, but with a
vanishing valence-to-conduction gap, $L$ being the length of the central segment. This pattern is the one expected
for usual semiconductors described by the (nonrelativistic) Schr\"{o}dinger equation, and it is associated with
zigzag nanoribbon central segments, irrespective of whether zigzag or metallic armchair leads are used.
\end{enumerate}

The faithful reproduction of these unique TB-NEGF conductance patterns by the DFP theory, including mixed 
armchair-zigzag configurations (where the carriers transit from a relativistic to a nonrelativistic regime),
provides a unifying framework for analysis of coherent transport phenomena and for interpretation of experiments
targeting fundamental understanding of transport in GNRs and the future development of graphene nanoelectronics.

\begin{figure*}
\centering\includegraphics[width=14cm]{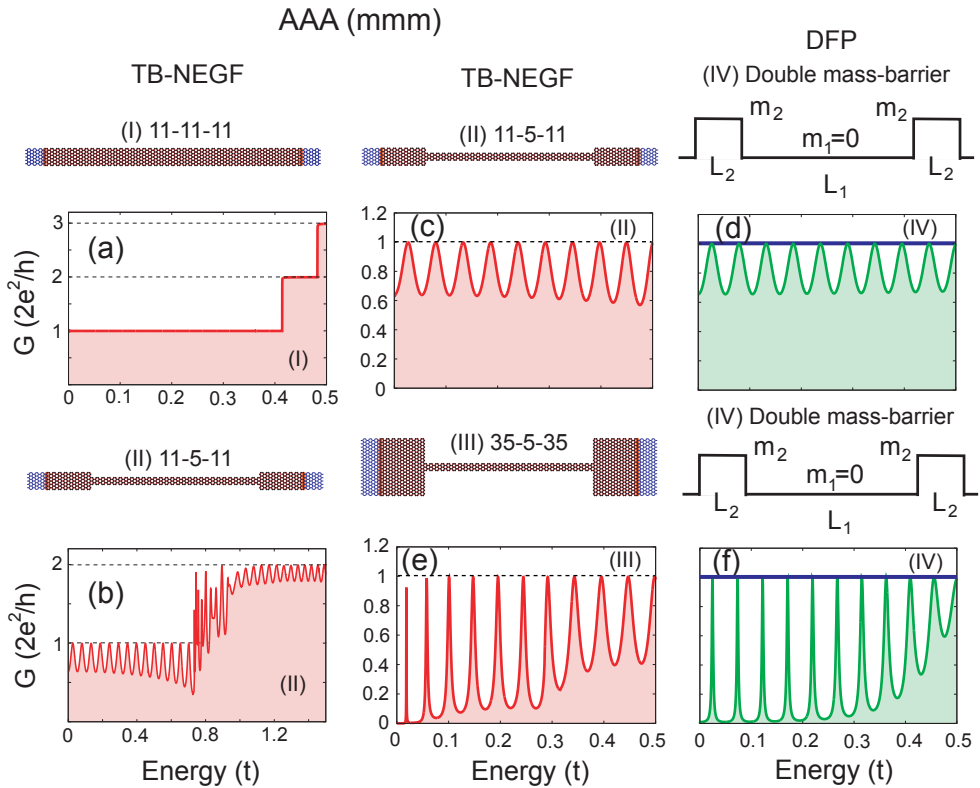}
\caption{{\bf Conductance quantization steps (a) for a uniform metallic armchair nanoribbon (I) contrasted to
Fabry-P\'{e}rot oscillations (b-f) for 3-segment all-metallic SaGNRs (II and III).}
The first two columns (employing red colors) display TB-NEGF results. The third column (employing 
green colors) displays continuum DFP results, which reproduce the TB-NEGF results in the middle
column. (I-III) Schematics of the nanoribbons employed in the TB-NEGF calculations. 
A 3-segment segmented GNR is denoted as ${\cal N}^W_1 - {\cal N}^W_2 - {\cal N}^W_1$, 
with ${\cal N}^W_i$ ($i=1,2$) being the number of carbon atoms specifying the width of the ribbon segments. 
In all configurations, the semi-infinite leads are shown in blue color on the far left and far right. 
Note the steps $nG_0$ in (a) reflecting full conductance quantization for the uniform nanoribbon. 
Instead of steps, the SaGNR junctions in (b-f) exhibit Fabry-P\'{e}rot-type conductance oscillations 
whose maxima are the places where conductance quantization is maintained. The effect of the relative widths 
of the leads and the constriction on the conductance of 
the all-metallic junctions is shown in (c,e) (TB-NEGF) and in (d,f) (corresponding DFP, respectively), illustrating 
Fabry-P\'{e}rot oscillations (c,d) for a constriction width close to that of the leads [see schematic (II)] 
and the development of a sharper oscillatory pattern [conductance spikes, see (e) and (f)] for a junction 
with much wider leads [see (III)]. The DFP approach is used to analyze the behavior of the TB-NEGF conductance in 
the energy range of the $1G_0$ step [see (b)]. (IV) Diagram of the double mass-barrier used in the DFP method 
[case of massless Dirac-Weyl electrons with $m_1=0$ and ${\cal M}_l=0$, ${\cal M}_l$ being the carrier mass in the
leads (not shown)]. The double-barrier parameters that reproduce the TB-NEGF results were 
$L_1=58 a_0$, $L_2=1 a_0$, $m_2 v_F^2=t/3$ for (d) and $L_1=60 a_0$, $L_2=6 a_0$, $m_2 v_F^2=t/4$ for (f). 
In the case of the wide leads (III), it is worth noting that the 
DFP theory reproduces the gradual widening of the spikes as a function of increasing energy; naturally this 
trend results from the weakening of the confinement effect due to a stronger coupling to the leads for 
higher energies. The horizontal solid lines at $G=1G_0$ (blue online) in (d) and (f) describe the DFP conductance 
obtained when employing a potential $V(x)$ double barrier [similar in shape to the schematic in (IV)] and the 
assumption $\phi(x)=0$; the result is independent of the potential barrier's heights.
In all figures (here and below), when a roman number is placed in the same frame along with a 
letter index, it indicates the corresponding lattice or DFP schematic specified by the roman number.
$a_0=0.246$ nm is the graphene lattice constant; $t=2.7$ eV is the hopping parameter.
}
\label{fig1}
\end{figure*}

~~~~~~~\\
{\bf Results}\\
~~~~~~~\\

{\bf Segmented Armchair GNRs: All-metallic}. 
Our findings for the case of all-metallic\cite{fuji96,waka10} segmented aGNRs (when the number of carbons 
specifying the width is ${\cal N}^W=3l+2$, $l=1,2,3,\ldots$) are presented in Fig.\ \ref{fig1}; this 
lattice configuration is denoted as AAA (mmm). A uniform metallic armchair GNR [see Fig.\ \ref{fig1}(I)] exhibits
ballistic quantized-conductance steps [see Fig.\ \ref{fig1}(a)]. In contrast, conductance quantization is absent
for a nonuniform 3-segment aGNR; see Figs.\ \ref{fig1}(b) $-$ \ref{fig1}(f). Instead of quantized steps, a finite
number of oscillations appears, whose maxima (at the value of unity) maintain a constant energy separation. This 
behavior is indicative of optical-like Fabry-P\'{e}rot multiple reflections of the DW electron within the cavity 
defined by the two contacts (interfaces between the segments of different width). The patterns in Figs.\ 
\ref{fig1}(b) $-$ \ref{fig1}(f) correspond to the category {\it FP-A1\/}. 

The dependence of the conductance on the width of the leads relative to that of the constriction is explored 
by comparing the two junctions (exhibiting sharp lead-to-constriction interfaces) depicted in Fig.\ 
\ref{fig1}(II) and Fig.\ \ref{fig1}(III). Examination of the TB-NEGF conductances for these two segmented
aGNRs [Fig.\ \ref{fig1}(c) and Fig.\ \ref{fig1}(e)] reveals that a wider lead [Fig.\ \ref{fig1}(III)] is 
associated with a stronger confinement (sharp conductance spikes) compared to a narrower lead [Fig.\ \ref{fig1}(II)] 
(oscillations). In the DFP results [Fig.\ \ref{fig1}(d) and Fig.\ \ref{fig1}(f)], which reproduce the 
TB-NEGF results, this trend is accounted for by varying the length and height of the mass barriers, which 
generates either a weak coupling (closed quantum-dot-like conductance spikes) or a strong coupling (open 
quantum-dot-like Fabry-P\'{e}rot-type oscillations) to the leads. 
     
Further insight can be gained by an analysis of the discrete energies associated with the humps 
of the conductance oscillations in Fig.\ \ref{fig1}(c) and the resonant spikes in Fig.\ 
\ref{fig1}(e). Indeed a simplified approximation for the electron confinement in the continuum model consists 
in considering the graphene electrons as being trapped within a 1D infinite-mass square well (IMSW) of 
length $L_1$ (the mass terms are infinite outside the interval $L_1$ and the coupling to the leads 
vanishes). The discrete spectrum of the electrons in this case is given \cite{fiol96} by
\begin{equation} 
E_n=\sqrt{\hbar ^2 v_F^2 k_n^2+{\cal M}^2 v_F^4},
\label{erel} 
\end{equation}
where the wave numbers $k_n$ are solutions of the 
transcendental equation
\begin{equation}
\tan(k_n L_1) = -\hbar k_n/({\cal M} v_F).
\label{eimsw}
\end{equation}  
In the context of Fig.\ \ref{fig1}, ${\cal M}=m_1=0$ (massless DW electrons), and one 
finds for the spectrum of the IMSW model: 
\begin{equation}
E_n=(n+1/2) \pi \hbar v_F/L_1,
\label{imsw0}
\end{equation} 
with $n=0,1,2,\ldots$. Remarkably, the energies associated with the TB-NEGF oscillation humps and spikes in 
Figs. \ref{fig1}(c) and \ref{fig1}(e) [or Figs.\ \ref{fig1}(d) and \ref{fig1}(f) in the DFP model] follow 
very closely the above relation. Note the constant separation between successive energies,
\begin{equation} 
E_n-E_{n-1}=2E_0, \;\;\; n=1,2,3,\ldots, 
\label{edis}
\end{equation}
which is twice the energy 
\begin{equation}
E_0=\pi \hbar v_F/(2L_1)
\label{eoff} 
\end{equation}
of the lowest state.
 
\begin{figure*}
\centering\includegraphics[width=17cm]{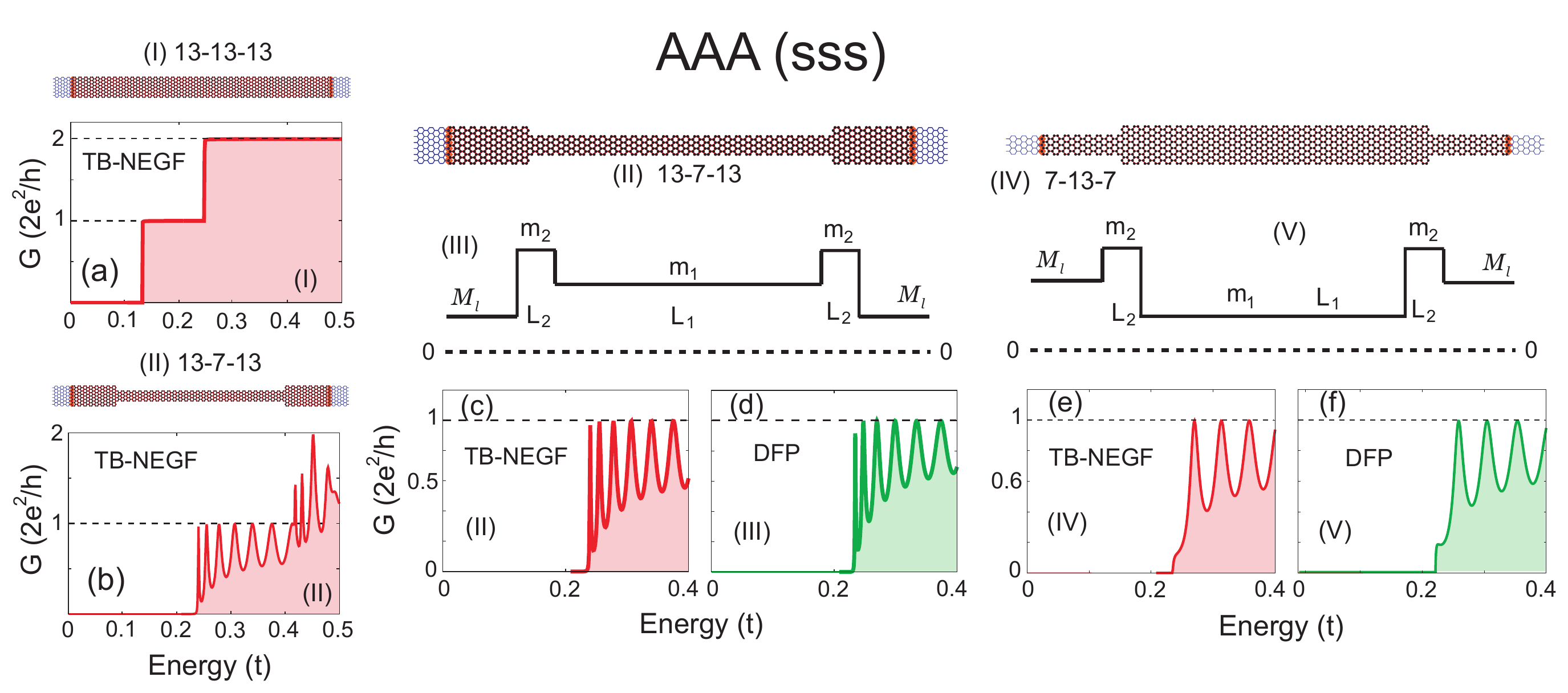}
\caption{{\bf Conductance quantization steps (a) for a uniform semiconducting armchair nanoribbon (I) contrasted 
to Fabry-P\'{e}rot oscillations (b-f) of two 3-segment armchair GNRs [(II) and (IV)] with both a semiconducting 
central constriction and semiconducting leads (13-7-13 and 7-13-7)}. (III, V) Schematics of the mass barriers 
used in the DFP modeling, with the dashed line denoting the zero mass. The physics underlying such a junction 
is that of a massive relativistic Dirac fermion impinging upon the junction and performing multiple 
reflections (above $m_1 v_F^2$) within a particle box defined by the double-mass barrier. 
(c,e) TB-NEGF conductance as a function of the Fermi energy of the {\it massive\/} Dirac electrons in the leads.
(d) DFP conductance reproducing [in the energy range of the $1G_0$ step, see (b)] the TB-NEGF result in (c).
The mass-barrier parameters used in the DFP reproduction were $L_1=55 a_0$,  $m_1 v_F^2= 0.22 t$, $L_2=1 a_0$,
$m_2 v_F^2=0.5 t$. The mass of the electrons in the leads was ${\cal M}_l v_F^2=0.166 t$.
(f) DFP conductance reproducing the TB-NEGF result in (e).
The parameters used in the DFP reproduction were $L_1=53.6 a_0$,  $m_1 v_F^2= 0.166 t$, $L_2=1 a_0$,
$m_2 v_F^2=0.51 t$. The mass of the electrons in the leads was ${\cal M}_l v_F^2=0.22 t$.
$a_0=0.246$ nm is the graphene lattice constant; $t=2.7$ eV is the hopping parameter.}
\label{fig2}
\end{figure*}

As is well known, a constant energy separation of the intensity peaks, inversely proportional to the length
of the resonating cavity [here $L_1$, see Eqs. (\ref{edis}) and (\ref{eoff}) above] is the hallmark of the 
optical Fabry-P\'{e}rot, reflecting the linear energy dispersion of the photon in optics or a massless DW 
electron in graphene structures. For our purpose, most revealing is the energy offset away form zero of the 
first conductance peak, which equals exactly one-half of the constant energy separation between the peaks. 
In onedimension, this is the hallmark of a massless fermion subject to an infinite- mass-barrier confinement, 
\cite{fiol96} and it provides ultimate support for our introduction of mass barriers at the interfaces of the 
segmented aGNR. Naturally, in the case of a semiconducting segment (see below), this equidistant 
behavior and $1/2$-offset of the conductance peaks do not apply; this case is accounted for by the 
Dirac-Fabry-P\'{e}rot model presented in Methods, and it is more general than the optical 
Fabry-P\'{e}rot theory associated with a photonic cavity.\cite{lipsbook} 

In the nonrelativistic limit, i.e., when $\hbar k_n \ll {\cal M} v_F$, one gets 
\begin{equation}
\tan(k_n L_1) \sim 0, 
\end{equation}
which yields the well known relations $k_n L_1 \sim n \pi$ and
\begin{equation} 
E_n \sim {\cal M}v_F^2 + n^2 \hbar^2 \pi^2/(2 {\cal M} L_1^2).
\label{ennr}
\end{equation}
For a massive relativistic electron, as is the case with the semiconducting aGNRs in this paper, one has to
numerically solve Eq.\ (\ref{eimsw}) and then substitute the corresponding value of $k_n$ in Einstein's energy 
relation given by Eq.\ (\ref{erel}). 

It is worth mentioning here that the inoperativeness for one-dimensional cases, due to Klein tunneling, \cite{kats06}
of the electrostatic potential $V(x)$ [see horizontal blue lines at $G=1G_0$ in Figs.\ \ref{fig1}(d) and \ref{fig1}(f)] 
was also noted earlier; see the curves for normal incidence (labeled $\theta=0$) in Fig.\ 2 of Ref.\ \onlinecite{rodr12}. 

\begin{figure}
\centering\includegraphics[width=7.5cm]{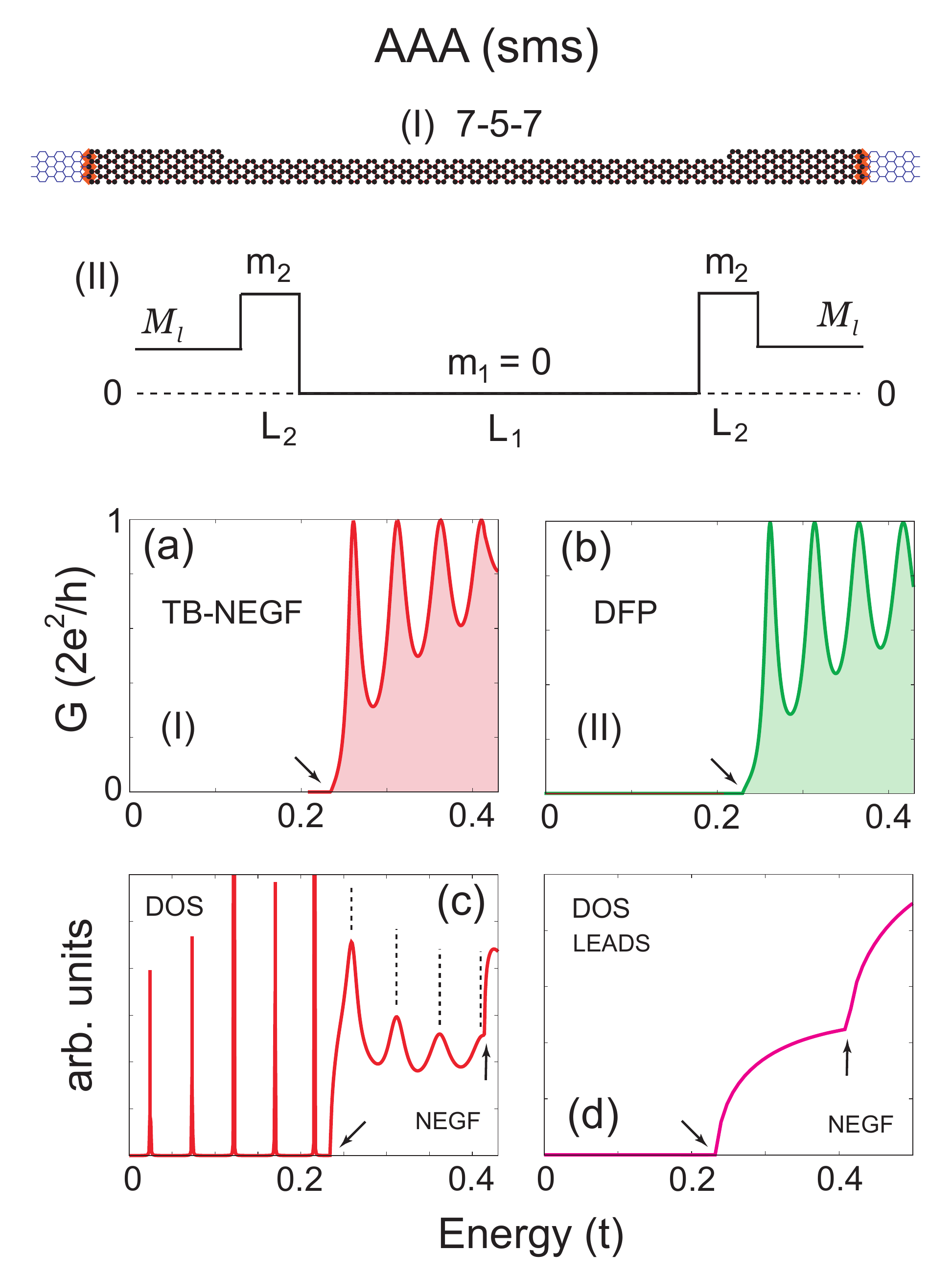}
\caption{{\bf Conductance for a 3-segment nanoribbon with a {\it metallic\/} (${\cal N}^W_2=5$) 
central constriction and semiconducting leads (${\cal N}^W_1=7$)}. The semi-infinite leads 
(in blue on the far left and far right) are an extension of the left and right semiconducting segments with 
${\cal N}^W_1=7$; see schematic lattice diagram in (I). (II) Schematics of the mass barriers 
used in the DFP modeling, with the dashed line denoting the zero mass. The physics underlying such a junction 
is that of a massive relativistic Dirac fermion impinging upon the junction, which loses its mass upon 
tunneling in the central segment and performs multiple reflections within a particle box defined by the 
double-mass barrier. (a) TB-NEGF conductance as a function of the Fermi energy of the {\it massive\/} Dirac 
electrons in the leads. (b) DFP conductance reproducing the TB-NEGF result in (a). The mass-barrier parameters
used in the DFP reproduction were $L_1=60.4 a_0$,  $m_1 = 0$, $L_2=1 a_0$, $m_2 v_F^2=0.37 t$. The mass of the 
electrons in the leads was ${\cal M}_l v_F^2=0.23 t$. (c)-(d)  The total DOS of the junction and the density
of states in the isolated leads, respectively, according to the TB-NEGF calculations. The arrows indicate the
onset of the electronic bands in the leads. Note that the DOS in (c) reveal the existence of five sharp
electronic states below the onset (at $0.23 t \equiv {\cal M}_l v_F^2$) of the first band in the leads [see
(d)], which consequently do not generate any conductance resonances [see (a) and (b)]. Note further in (c) the 
equal energy spacing between the vertical lines [the five solid (red) and four dashed (black) ones] associated
with the resonances of a {\it massless\/} electron confined within the central metallic aGNR segment.   
$a_0=0.246$ nm is the graphene lattice constant; $t=2.7$ eV is the hopping parameter.}
\label{fig3}
\end{figure}

{\bf Segmented Armchair GNRs: All-semiconducting}.
Our results for a 3-segment {\it all-semiconducting\/} aGNR are portrayed in Fig.\ \ref{fig2} [see schematic 
lattice diagrams in Figs.\ \ref{fig2}(I) and \ref{fig2}(II)]; this lattice configuration is denoted as AAA (sss). 
A uniform semiconducting armchair GNR [see Fig.\ \ref{fig2}(I)] exhibits ballistic quantized-conductance steps 
[see Fig.\ \ref{fig2}(a)]. In contrast, conductance quantization is absent for a nonuniform 3-segment ($13-7-13$)
aGNR; see Figs.\ \ref{fig2}(b) $-$ \ref{fig2}(d). Here, instead of quantized steps, 
oscillations appear as in the case of
the all-metallic junctions presented earlier in Fig.\ \ref{fig1}. However, the first oscillation appears now
at an energy $\sim 0.22 t$, which reflects the intrinsic gap $\Delta/2$ of the semiconducting central segment 
belonging to the class II of aGNRs, specified\cite{waka10,yann14} by a width ${\cal N}^W=3l+1$, 
$l=1,2,3,\ldots$. That the leads are semiconducting does not have any major effect.
This is due to the fact that ${\cal N}^W_2 < {\cal N}^W_1$, and as a result the energy gap $m_1 v_F^2$ of the 
central segment is larger than the energy gap ${\cal M}_l v_F^2$ of the semiconducting leads [see schematic
in Fig.\ \ref{fig2}(III)]. 

The armchair GNR case with interchanged widths (i.e., $7-13-7$ instead of $13-7-13$) is portrayed in Figs.\ 
\ref{fig2}(e) $-$ \ref{fig2}(f). In this case the energy gap of the semiconducting leads (being the largest) 
determines the onset of the conductance oscillations. It is a testimonial of the consistency of our DFP method that
it can reproduce [see Fig.\ \ref{fig2}(d) and Fig.\ \ref{fig2}(f)] both the $13-7-13$ and 
$7-13-7$ TB-NEGF conductances; this is achieved with very similar sets of parameters taking into consideration the
central-segment-leads interchange. We note that the larger spacing between peaks (and also the smaller number of
peaks) in the $7-13-7$ case is due to the smaller mass of the central segment ($0.166t$ instead of $0.22t$). 

From an inspection of Fig.\ \ref{fig2}, one can conclude 
that the physics associated with the all-semiconducting AAA junction is that of multiple reflections of a 
{\it massive\/} relativistic Dirac fermion bouncing back and forth from the edges of a particle box created by 
a double-mass barrier [see the schematic of the double-mass barrier in Fig.\ \ref{fig2}(III)]. In particular, 
to a good approximation the energies of the conductance oscillation peaks are given by the IMSW Eq.\
(\ref{eimsw}) with ${\cal M}v_F^2=m_1 v_F^2=0.22 t$ ($13-7-13$) or ${\cal M}v_F^2=m_1 v_F^2=0.166 t$ ($7-13-7$). 
In this respect, the separation energy between successive peaks in Figs.\ \ref{fig2}(b), \ref{fig2}(c),
\ref{fig2}(e), and \ref{fig2}(f) is not a constant, unlike the case of the all-metallic junction. 

The patterns in Figs.\ \ref{fig2}(c) and \ref{fig2}(f) correspond to the category {\it FP-B\/}. 
This generalized oscillations cannot be accounted for by the optical Fabry-P\'{e}rot theory, but they are well
reproduced by the generalized Dirac-Fabry-P\'{e}rot model introduced by us in the Methods.

{\bf Segmented Armchair GNRs: semiconducting-metallic-semiconducting}.
Our results for a 3-segment ($7-5-7$) {\it semiconducting-metallic-semiconducting\/} aGNR are portrayed in Fig.\ 
\ref{fig3} [see schematic lattice diagram in Fig.\ \ref{fig3}(I)]; this lattice configuration is denoted as AAA (sms). 
The first FP oscillation in the TB-NEGF conductance displayed in Fig.\ \ref{fig3}(a) appears at an energy 
$\sim 0.23 t$, which reflects the intrinsic gap $\Delta/2$ of the semiconducting leads (with ${\cal N}^W_1=7$). The
energy spacing between the peaks in Fig.\ \ref{fig3}(a) is constant in agreement with the metallic (massless DW
electrons) character of the central segment with ${\cal N}^W_2=5$. The TB-NEGF pattern in Fig.\ \ref{fig3}(a)
corresponds to the Fabry-P\'{e}rot category {\it FP-A2\/}. As seen from Fig.\ \ref{fig3}(b), our generalized 
Dirac-Fabry-P\'{e}rot theory is again capable of faithfully reproducing this behavior.

A deeper understanding of the AAA (sms) case can be gained via an inspection of the density of states (DOS) plotted in
Fig.\ \ref{fig3}(c) for the total segmented aGNR (central segment plus leads) and in Fig.\ \ref{fig3}(d) for the  
the isolated leads. In Fig.\ \ref{fig3}(c), nine equidistant resonance lines are seen. Their energies are close to
those resulting from the IMSW Eq.\ (\ref{imsw0}) (with $L_1=60.4 a_0$, see the caption of Fig.\ \ref{fig3}) for a
massless DW electron. Out of these nine resonances, the first five do not conduct [compare Figs.\ \ref{fig3}(a) and
\ref{fig3}(c)] because their energies are lower than the minimum energy (i.e., $\Delta/2={\cal M}_l v_F^2 \sim
0.23 t$) of the incoming electrons in the leads [see the onset of the first band (marked by an arrow) in the
DOS curve displayed in Fig.\ \ref{fig3}(d)]. 

\begin{figure}
\centering\includegraphics[width=7.5cm]{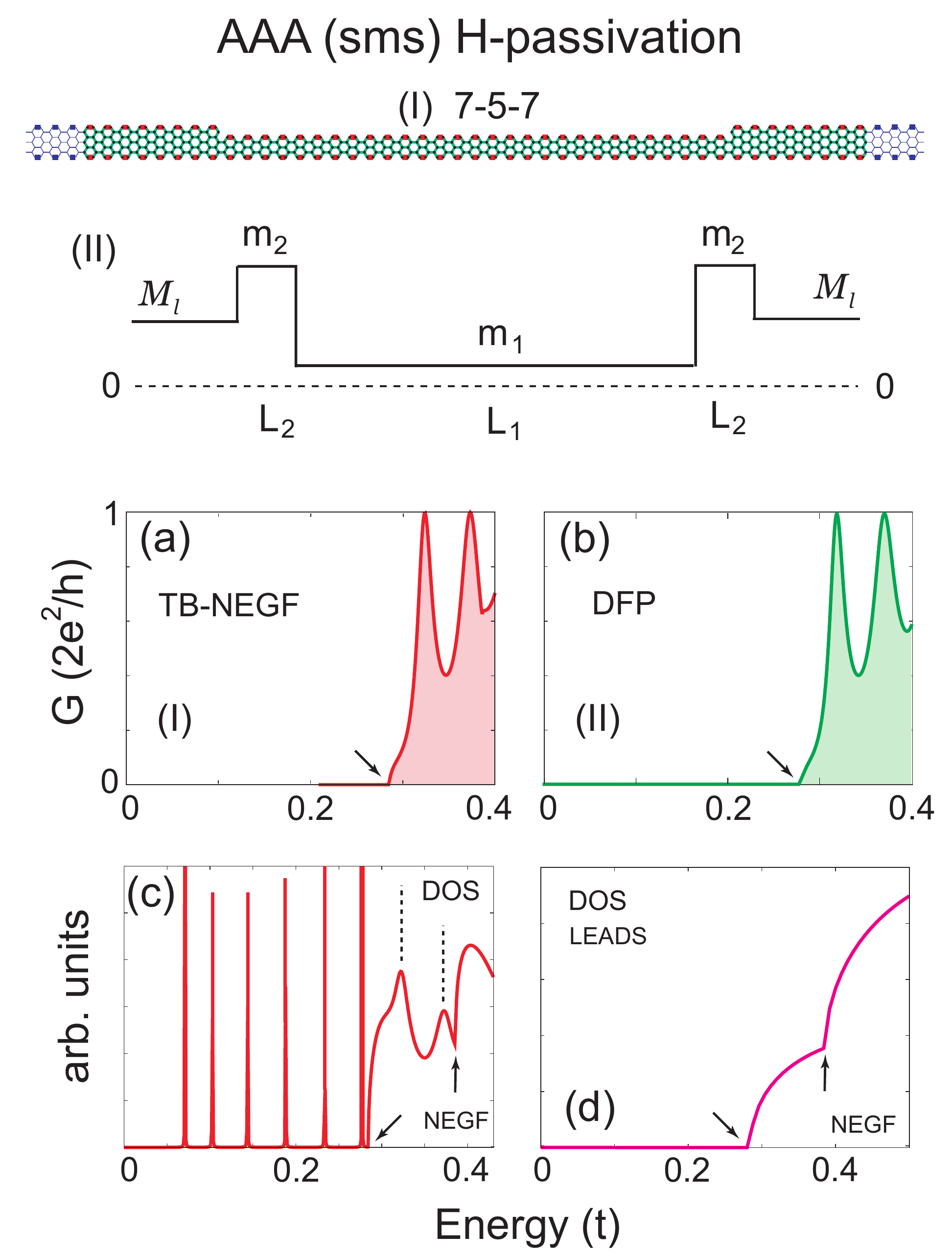}
\caption{{\bf H-passivation effects in the conductance of a 3-segment armchair nanoribbon with a {\it metallic\/} 
(${\cal N}^W_2=5$) central constriction and semiconducting leads (${\cal N}^W_1=7$)}; see schematic 
lattice diagram in (I). Note that the nearest-neighbor C-C bonds at the armchair edges 
(thick red and blue lines) have hopping parameters $t^\prime=1.12 t$. (II) Schematics of the
position-dependent mass field used in the DFP modeling, with the dashed line denoting the zero mass. The
physics underlying such a junction 
is that of a massive relativistic Dirac fermion impinging upon the junction, which reduces its mass close to
zero upon tunneling in the central segment and performs multiple reflections within a particle box defined by the 
double-mass barrier. (a) TB-NEGF conductance as a function of the Fermi energy. (b) DFP conductance reproducing the 
TB-NEGF result in (a). The mass parameters used in the DFP reproduction were $L_1=59.5 a_0$,  $m_1 v_F^2= 0.05 t$, 
$L_2=1.5 a_0$, $m_2 v_F^2=0.30 t$. The mass of the electrons in the leads was ${\cal M}_l v_F^2=0.28t$. (c)-(d) 
The total DOS of the junction and the density of states in the isolated leads, respectively, according to the TB-NEGF
calculations. The arrows indicate the onset of the electronic bands in the leads; note the shifts from $0.23 t$ to
$0.28 t$ and from $0.42t$ to $0.38t$ for the onsets of the first and second bands, respectively, compared to the
case with $t^\prime=t$ in Fig.\ \ref{fig3}(d). Compared to Fig.\ \ref{fig3}, the subtle modifications of mass 
parameters brought about by having $t^\prime =1.12 t$ result in having six sharp electronic states [see (c)] below the
onset (at $0.28 t \equiv {\cal M}_l v_F^2$) of the first band in the leads [see (d)], which consequently do not 
generate any conductance resonances [see (a) and (b)]. In addition, within the energy range ($0$ to $0.4t$) shown in
(a) and (b), there are now only two conducting resonances, instead of three compared to Figs.\ \ref{fig3}(a) and
\ref{fig3}(b). $a_0=0.246$ nm is the graphene lattice constant; $t=2.7$ eV is the graphene hopping parameter.}
\label{fig4}
\end{figure}

{\bf Segmented Armchair GNRs: Effects of hydrogen passivation}.
As shown in Refs.\ \onlinecite{cohe06,wang07}, a detailed description of hygrogen passivation requires that the
hopping parameters $t^\prime$ for the nearest-neighbor C-C bonds at the armchair edges be given by $t^\prime=1.12t$.   
Taking this modification into account, our results for a 3-segment {\it semiconducting-metallic-semiconducting\/} 
aGNR are portrayed in Fig.\ \ref{fig4} [see schematic lattice diagram in Fig.\ \ref{fig4}(I)]; 
this lattice configuration is denoted as "AAA (sms) H-passivation." The first FP oscillation in the 
TB-NEGF conductance displayed 
in Fig.\ \ref{fig4}(a) appears at an energy $\sim 0.28 t$, which reflects the intrinsic gap $\Delta/2$ of the 
properly passivated semiconducting leads (with ${\cal N}^W_1=7$). The energy spacing between the peaks in Fig.\ 
\ref{fig4}(a) is slightly away from being constant in agreement with the small mass $m_1 v_F^2=0.05t$ acquired by
the central segment with ${\cal N}^W_2=5$, due to taking $t^\prime=1.12t$. As seen from Fig.\ \ref{fig4}(b), our
generalized Dirac-Fabry-P\'{e}rot theory is again capable of faithfully reproducing this behavior.

A deeper understanding of the AAA (sms)-H-passivation case can be gained via an inspection of the DOS plotted in 
Fig.\ \ref{fig4}(c) for the total segmented aGNR (central segment plus leads) and in Fig.\ \ref{fig4}(d) for the 
isolated leads. In Fig.\ \ref{fig4}(c), eight (almost, but not exacrly, equidistant) resonance lines are
seen. Their energies are close to those resulting from the IMSW Eq.\ (\ref{eimsw}) 
(with $L_1=59.5 a_0$ and $m_1 v_F^2=0.05t$; see the
caption of Fig.\ \ref{fig4}) for a Dirac electron with a small mass. Out of these eight resonances, the first 
six do not conduct [compare Figs.\ \ref{fig4}(a) and \ref{fig4}(c)] because their energies are lower than the minimum energy
(i.e., $\Delta/2={\cal M}_l v_F^2 \sim 0.28 t$) of the incoming electrons in the leads [see the onset of the first 
band (marked by an arrow) in the DOS curve displayed in Fig.\ \ref{fig4}(d)]. 
From the above we conclude that hydrogen passivation of the aGNR resulted in a small shift of the location of
the states, and opening of a small gap for the central metallic narrower (with a width of $N_2^W=5$) segment, but did
not modify the conductance record in any qualitative way. Moreover, the passivation effect can be faithfully
captured by the Dirac FP model by a small readjustment of the model parameters. 

\begin{figure}
\centering\includegraphics[width=7.5cm]{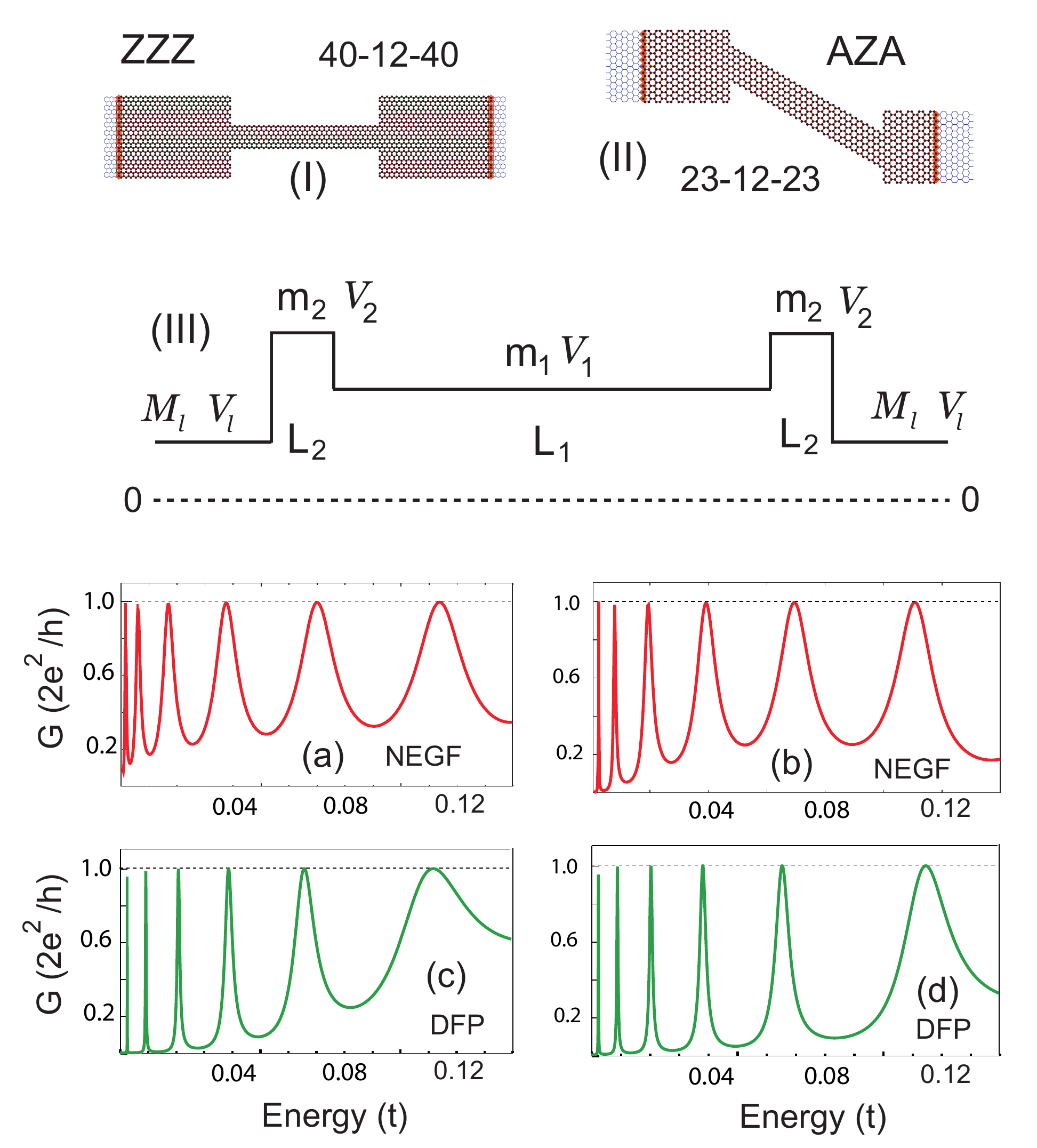}
\caption{{\bf Conductance for ZZZ (all-zigzag, left column) and AZA (armchair-zigzag-armchair, right column) 
segmented nanoribbon junctions.} See corresponding lattice diagrams in (I) and (II). 
The 3-segment GNRs are denoted as ${\cal N}^W_1 - {\cal N}^W_2 - {\cal N}^W_1$, with ${\cal N}^W_i$ ($i=1,2$) 
being the number of carbon atoms specifying the width of the ribbon segments. The armchair leads in 
the AZA junction are metallic (${\cal N}^W_1=23$, class III aGNR).  
(a)-(b) TB-NEGF conductance for the ZZZ and AZA junction, respectively. 
(c) DFP conductance reproducing the TB-NEGF result in (a) for the ZZZ junction.
(d) DFP conductance reproducing the TB-NEGF result in (b) for the AZA junction.  
In spite of the different edge morphology, the Fabry-P\'{e}rot patterns in (a) and (b) are very similar.
The central zigzag segment controls the Fabry-P\'{e}rot patterns. According to the continuum DFP analysis,
the physics underlying such patterns is that of a massive nonrelativistic Schr\"{o}dinger fermionic carrier 
performing multiple reflections within a cavity defined by a double-mass barrier [see diagram in (III)], but 
with the additional feature that $V_1=-m_1 v_F^2$ and $V_l=-{\cal M}_l v_F^2$ are also considered for segments
or leads with zigzag edge terminations (see text for details). The mass and $V_i$ parameters used in the DFP 
calculations were $L_1=30 a_0$,  $m_1 v_F^2=2.23 t - c E t$, with $c=7.3$, $V_1=-m_1 v_F^2$, $L_2=1.1 a_0$, 
$m_2 v_F^2=0.38 t$, $V_2=-m_2 v_F^2/3$, ${\cal M}_l v_F^2= 2.30 t$, $V_l=-{\cal M}_l v_F^2$ in (c) and 
$L_1=29.1 a_0$,  $m_1 v_F^2=2.65 t - c E t$, with $c=8.4$, $V_1=-m_1 v_F^2$, $L_2=1.0 a_0$, $m_2 v_F^2=0.30 t$, 
$V_2=-m_2 v_F^2$, ${\cal M}_l=0$, $V_l=0$ in (d). ${\cal M}_l$ and $V_l$ denote parameters of the leads.
$E$ is the energy in units of $t$. $a_0=0.246$ nm is the graphene lattice constant; $t=2.7$ eV is the hopping 
parameter.}
\label{fig5}
\end{figure}

{\bf All-zigzag segmented GNRs}.
It is interesting to investigate the sensitivity of the interference features on the edge morphology.
We show in this section that the relativistic transport treatment applied to segmented armchaie GNRs does not 
maintain for the case of a nanoribbon segment with zigzag edge terminations. In fact zigzag GNR (zGNR) segments
exhibit properties akin to the well-known transport in usual semiconductors, i.e., their 
excitations are governed by the nonrelativistic Schr\"{o}dinger equation.

Before discussing segmented GNRs with zigzag edge terminations, we remark that such GNRs with uniform width
exhibit stepwise quantization of the conductance, similar to the case of a uniform metallic 
armchair-edge-terminated GNR [see Fig.\ \ref{fig2}(a)].

In Fig.\ \ref{fig5}(a), we display the conductance in a three-segment junction [see lattice schematic in Fig.\ 
\ref{fig5}(I)] when all three segments have zigzag edge terminations (denoted as ZZZ).  
The main finding is that the central segment behaves again as a resonant cavity that yields 
an oscillatory conductance pattern where the peak spacings are unequal [Fig.\ \ref{fig5}(a)]. 
This feature, which deviates from the optical Fabry-P\'{e}rot behavior, appeared also in the DFP patterns for
a three-segment armchair junction whose central segment was semiconducting, albeit with a different dependence
on $L$ [see Figs.\ \ref{fig2} and Eq.\ (\ref{eimsw})]. Moreover, from a set of
systematic calculations (not shown) employing different lengths and widths, we found that the energy of
the resonant levels in zGNR segments varies on the average as $\sim (n/L)^2$, where the integer $n$ counts the 
resonances and $L$ indicates the length of the central segment. However, a determining difference with
the armchair GNR case in Fig.\ \ref{fig2} is the vanishing of the valence-to-conductance gap
in the zigzag case of Fig.\ \ref{fig5}(a). It is well known that the above features 
are associated with resonant transport of electronic excitations that obey the second-order 
nonrelativistic Schr\"{o}dinger equation.

Naturally, one could formulate a continuum transport theory based on transfer matrices (see Methods) 
that use the 1D Schr\"{o}dinger equation instead of the generalized Dirac Eq.\ (\ref{direq}).  
Such a Schr\"{o}dinger-equation continuum approach, however, is unable to describe mixed armchair-zigzag
interfaces (see below), where the electron transits between two extreme regimes, i.e., an 
ultrarelativistic (i.e., including the limit of vanishing carrier mass) Dirac regime (armchair segment) and a 
nonrelativistic Schr\"{o}dinger regime (zigzag segment). We have thus been led to adopt the same Dirac-type 
transfer-matrix approach as with the armchair GNRs, but with nonvanishing potentials $V=\eta {\cal M} v_F^2$,
with $\eta=\Theta(-E)-\Theta(E)$, where $\Theta(E)$ is the Heaviside step function. 
This amounts to shifts (in opposite senses) of the energy scales for particle and hole excitations, respectively,
and it yields the desired vanishing value for the valence-to-conduction gap of a zigzag GNR.

The calculated DFP conductance that reproduces well the TB-NEGF result for the ZZZ junction [Fig.\
\ref{fig5}(a)] is displayed in Fig.\ \ref{fig5}(c); the parameters used in the DFP calculation 
are given in the caption of Fig.\ \ref{fig5}. We note that the carrier mass ($m_1$) in the central zigzag segment 
exhibits an energy dependence. This is similar to a well known effect (due to nonparabolicity in the $E-k$ 
dispersion) in the transport theory of usual semiconductors.\cite{melc93} We further note that the average 
mass associated with a zigzag segment is an order of magnitude larger than that found for semiconducting 
armchair segments of similar width (see captions in Fig.\ \ref{fig2}, and this yields energy 
levels $\sim (n/L)^2$ close to the nonrelativistic limit [see Eq.\ (\ref{ennr})]. We note that FP pattern of the ZZZ
junction belongs to the category {\it FP-C\/}. 

{\bf Mixed armchair-zigzag-armchair segmented GNRs}.
Fig.\ \ref{fig5} (right column) presents an example of a mixed armchair-zigzag-armchair (AZA) junction, where 
the central segment has again zigzag edge terminations [see lattice schematic in Fig.\ \ref{fig5}(II)]. The 
corresponding TB-NEGF conductance is displayed in Fig.\ \ref{fig5}(b). In spite of the different morphology 
of the edges between the leads (armchair) and the central segment (zigzag), the conductance profile of
the AZA junction [Fig.\ \ref{fig5}(b)] is very similar to that of the ZZZ junction [Fig.\ \ref{fig5}(a)].
This means that the characteristics of the transport are determined mainly by the central segment, with the
left and right leads, whether zigzag or armchair, acting as reservoirs supplying the impinging electrons. 

The DFP result reproducing the TB-NEGF conductance is displayed in Fig.\ \ref{fig5}(d), and the parameters
used are given in the caption. We stress that the mixed AZA junction represents a rather unusual physical
regime, where an ultrarelativistic Dirac-Weyl massless electron (due to the metallic armchair GNRs in the leads) 
transits to a nonrelativistic massive Schr\"{o}dinger one in the central segment. We note that FP pattern of the 
AZA junction belongs to the category {\it FP-C\/}. 

\begin{figure}
\centering\includegraphics[width=7.5cm]{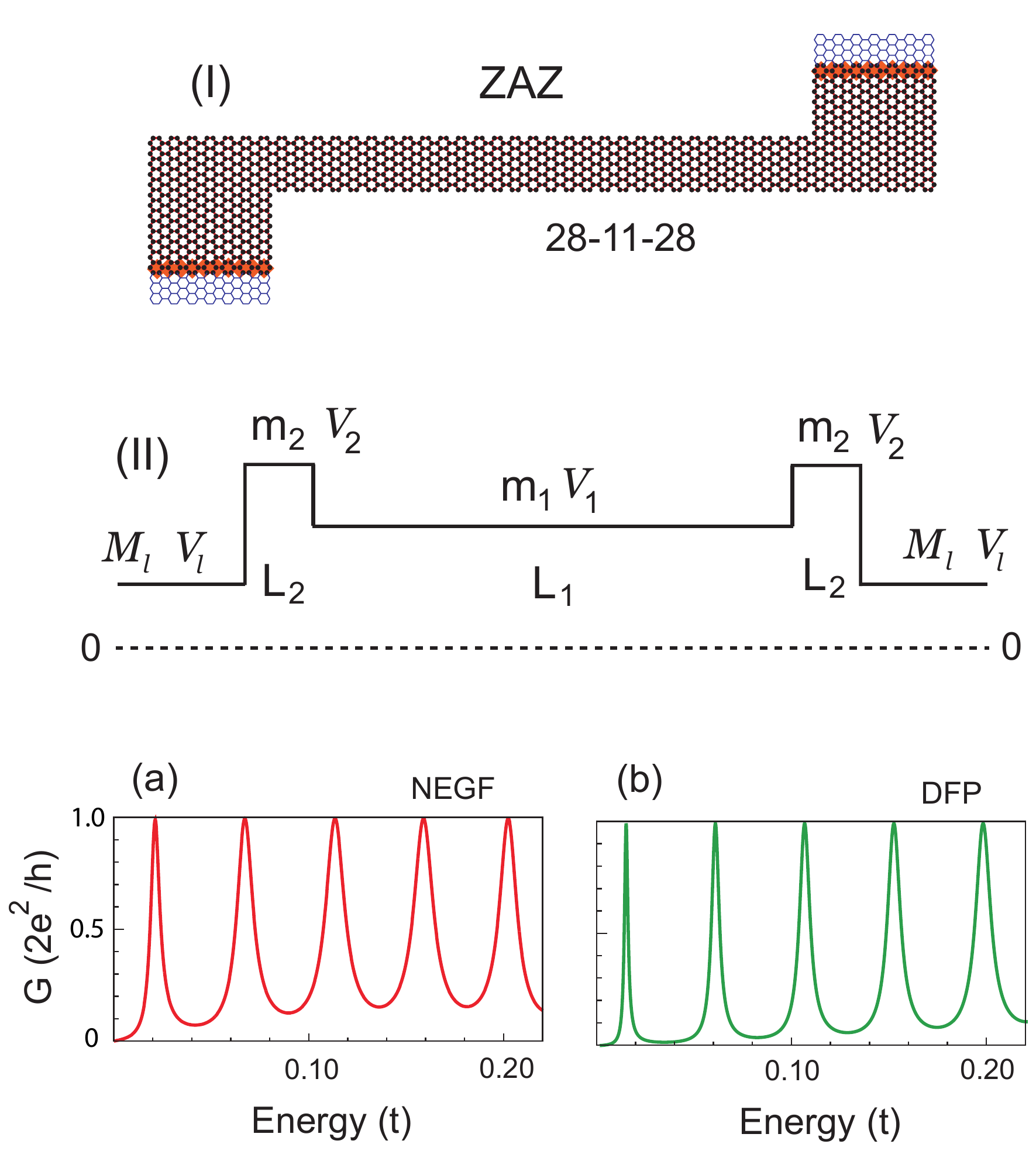}
\caption{{\bf Conductance for a ZAZ (zigzag-armchair-zigzag) segmented nanoribbon junction.} 
See corresponding lattice diagram in (I).
The 3-segment GNRs are denoted as ${\cal N}^W_1 - {\cal N}^W_2 - {\cal N}^W_1$, with ${\cal N}^W_i$ ($i=1,2$) 
being the number of carbon atoms specifying the width of the ribbon segments. The central armchair segment in 
the ZAZ junction is metallic (${\cal N}^W_2=11$, class III aGNR).  
(a) TB-NEGF conductance for the ZAZ junction. 
(b) DFP conductance reproducing the TB-NEGF result in (a) for the ZAZ junction.
The central armchair segment controls the Fabry-P\'{e}rot patterns. According to the continuum DFP analysis,
the physics underlying such patterns is that of a massless relativistic Dirac-Weyl fermionic carrier 
performing multiple reflections within a cavity defined by a double-mass barrier [see diagram in (II)]. 
but with the additional feature that $V_l=-{\cal M}_l v_F^2$ are also considered for the
leads with zigzag edge terminations (see text for details). The mass and $V_i$ parameters used in the DFP 
calculations were $L_1=66 a_0$, $m_1=0$, $V_1=0$, $L_2=1 a_0$, $m_2 v_F^2=0.50 t$, $V_2=0$, 
${\cal M}_l v_F^2=2 t$. $V_l=-{\cal M}_l v_F^2$.
$E$ is the energy in units of $t$. $a_0=0.246$ nm is the graphene lattice constant; $t=2.7$ eV is the hopping 
parameter.}
\label{fig6}
\end{figure}

{\bf Mixed zigzag-armchair-zigzag segmented GNRs}.
Finally Fig.\ \ref{fig6} presents an example of a mixed zigzag-armchair-zigzag (ZAZ) junction, where 
the central segment corresponds to a metallic armchair GNR [see lattice schematic in Fig.\ \ref{fig6}(I)]. The 
corresponding TB-NEGF conductance is displayed in Fig.\ \ref{fig6}(a). In spite of the different morphology 
of the edges between the leads (zigzag) and the central segment (armchair), the conductance profile of
the ZAZ junction [Fig.\ \ref{fig6}(a)] is controlled by determined the central segment, with the
left and right leads acting as reservoirs supplying the impinging electrons. Thus the conductance peaks are
close to being equidistant, and the FP pattern belongs to the categoty {\it FP-A1\/}. 
The DFP result reproducing the TB-NEGF conductance is displayed in Fig.\ \ref{fig6}(b), and the parameters
used are given in the caption. 

In a reverse sense compared to the AZA junction above, the mixed ZAZ junction here represents also 
a rather unique physical regime, where a nonrelativistic massive Schr\"{o}dinger electron (due to the
zigzag GNRs in the leads) transits to an ultrarelativistic Dirac-Weyl massless one in the central segment. The 
decisive advances brought forward by our DFP 1D continuum theory can be clearly appreciated by its ability to 
describe the corresponding TB-NEGF results for these highly nontrivial ZAZ and AZA junctions [compare Figs.\ 
\ref{fig6}(a) and \ref{fig6}(b), as well as Figs.\ \ref{fig5}(b) and \ref{fig5}(d)]. 
Contimuum 2D formulations are unable to describe the
important all-zigzag and mixed armchair/zigzag junctions described in this section, because they 
cannot distinguish between armchair and zigzag edges.  

~~~~~~\\
{\bf Discussion}\\
~~~~~~~~\\

To motivate the 1D Dirac formalism developed and utilized here, we note that 
customarily armchair or zigzag edge terminations in two-dimensional (2D) graphene nanoribbons are treated
with the massless 2D Dirac equation with the use of the corresponding boundary conditions. \cite{brey06} In our study,
however, narrow GNRs are treated with a 1D generalized Dirac equation. This approach is reminiscent of the 1D 
description of GNRs as described in Ref.\ \onlinecite{zhen07} and also reviewed in Ref.\ \onlinecite{waka10}, where 
the tight-binding 2D spectrum is projected onto the longitudinal wavenumber direction, $k_x$. For the
armchair case (see Eqs.\ 10 on Ref.\ \onlinecite{zhen07}), the low-energy range of
the 1D spectrum can be well approximated by the Einstein energy relation with a non-zero mass term for the
semiconducting case, and with a zero mass term for the metallic case which exhibits a massless linear dependence
(photon-like dispersion) of the energy on the momentum. 

As a function of $k_x$, the zigzag GNRs exhibit a partially flat $E \sim 0$ band due to the localized (in the 
transverse direction) edge state. The flat part is centered around $k_x = \pi$ (alternatively taken as $k_x=0$
in Ref.\ \onlinecite{brey06}), where $E(k_x=\pi)=0$, and it expands towards the graphene Dirac points located
at $2\pi/3$ (alternatively, $k_x = -\pi/3$) and $4\pi/3$ (alternatively $k_x = \pi/3$) as the width of the
ribbon increases (see Fig. 7 in Ref.\ \onlinecite{waka10}). After reaching the ends of the flat part, the
energy band opens two branches with non-zero energies; these branches do not have a linear dependence
on $k_x$. Although this can be termed also as a ``gapless'' spectrum (like the metallic armchair case), the
two differ in their dispersion relation -- that is the energy of the former (zigzag) is approximately
independent of the  momentum for a broad range of the $k_x$ momentum (see examples for various widths of
zigzag GNRs in Fig.\ 4 of Ref.\ \onlinecite{waka10}), whereas the latter (armchair) shows a linear
energy-momentum relationship (see Fig.\ 3 in Ref.\ \onlinecite{waka10}). 

As the width of the GNRs get smaller (narrow zigzag nanoribbons), the range of the aforementioned flat part decreases,
and in the limit of a single row of benzene rings (the narrow-most zigzag GNR, referred to as polyacene) the dispersion
curve transforms \cite{kive83} into two (electron-hole) parabolas touching at $k_x=\pi$. The spectrum of
polyacene is gapless, but due to the nonrelativistic parabolic dispersion, a mass, ${\cal M}$, is associated 
\cite{kive83} with the second derivative according to the nonrelativistic relation $p^2/2{\cal M}$. The case
of narrow zigzag GNRs that we study here is closer to the nonrelativistic polyacene than to the relativistic
massless graphene (i.e., the limit of a zigzag ribbon of infinite width). Incidentally, we mention that the
polyacene parabolic-band limit is not obtained \cite{onip08} through a treatment employing the massless 2D
Dirac equation with boundary conditions. 

We comment here that physical circumstances where a change in some system configuration can result in a change
in the nature of the dispersion relation (e.g  from linear to quadratic), are not that rare. Another example
is found when comparing the gapless, but linear in momentum, dispersion of the energy of particles (massless 
relativistic) in a monolayer 2D graphene sheet with the gapless, but parabolic, dispersion (massive
non-relativistic particles) in bilayer graphene.\cite{mcca13}

Furthermore, our NEGF conductance calculations for segmented zigzag GNRs exhibited Fabry-P\'{e}rot
oscillatory patterns, with the spacing between peaks behaving as $1/L^2$ (where $L$ is the length of the GNR segment).
This differs from the (photon-like) FP pattern found for the metallic case of armchair GNRs where the
spacing between peaks in the oscillatory conductance varies as $1/L$ (characteristic of {\it massless\/} particles with
linear energy-momentum dispersion, e.g. photons), which were studied \cite{lipsbook} by Fabry and P\'{e}rot. This
directly suggests that the zigzag GNRs can be described by a nonrelativistic limit of the Dirac equation with a
sufficiently large mass (i.e. by the Schr\"{o}dinger equation).

Focusing on the intrinsic properties of the graphene lattice, originating from the topology
of the honeycomb network, and using a NEGF approach, we have studied here the transport properties of 
atomically precise segmented armchair and zigzag graphene nanoribbons, with segments of different widths. 
Mixed armchair-zigzag junctions (with segments of different widths) have also been discussed. 

The electronic conductance is found to exhibit Fabry-P\'{e}rot oscillations, or resonant tunneling, associated 
with partial confinement and formation of a quantum box (resonant cavity) in the junction. The Fabry-P\'{e}rot oscillations occur for junctions that are strongly coupled to the leads (open system), whereas the resonant-tunneling
spikes appear for weak lead-junction coupling (closed system). In particular, with regard to the FP interference
patterns, three distinct categories were identified (see the Introduction), with only one of them having the
characteristics of the optical \cite{lipsbook} FP pattern corresponding to {\it massless\/} graphene electrons 
exhibiting equal spacings between neighboring peaks.

Perfect quantized-conductance flat steps were found only for uniform GNRs. 
In the absence of extraneous factors, like disorder, in our theoretical model, the 
deviations from the perfect quantized-conductance steps were unexpected. However, this aforementioned behavior
obtained through TB-NEGF calculations is well accounted for by a 1D contimuum fermionic Dirac-Fabry-P\'{e}rot 
interference theory (see Methods). This approach employs an effective position-dependent mass term in the 
Dirac Hamiltonian to absorb the finite-width (valence-to-conduction) gap in armchair nanoribbon segments, 
as well as the barriers at the interfaces between nanoribbon segments forming a junction. For zigzag
nanoribbon segments the mass term in the Dirac equation reflects the nonrelativistic Schr\"{o}dinger-type 
behavior of the excitations. We emphasize that the mass in zigzag-terminated GNR segments is much larger 
than the mass in semiconducting armchair-terminated GNR
segments. Furthermore in the zigzag GNR segments (which are always characterized by a vanishing 
valence-to-conduction energy gap), the mass corresponds simply to the carrier mass. In the armchair GNR
segments, the carrier mass endows (in addition) the segment with a valence-to-conduction energy gap,
according to Einstein's relativistic energy relation [see Eq.\ (\ref{erel})].

We observe here that the Dirac Fabry-P\'{e}rot masses that we find to yield agreement with the TB conductance
spectra agree well with those obtained through Density Functional Theory (DFT) calculations \cite{cohe06}
where the energy gap $\Delta/2={\cal M} v_F^2$ for armchair GNRs versus width has been determined. For example,
for ${\cal N}^W=5$ and ${\cal N}^W=7$, Eq.\ (1) in Ref.\ \onlinecite{cohe06} yields $\Delta/2=0.06t$ and $0.28t$,
respectively. These DFT values agree well with the DFP values of $0.05t$ (central segment) and $0.28t$ (leads) given
in the caption of Fig.\ \ref{fig4}, where the H-passivation effect according to the DFT was incorporated in the TB
description. 

The above findings point to a most fundamental underlying physics, namely that the topology of disruptions of
the regular  honeycomb lattice (e.g., variable width segments, corners, edges) generate a 
scalar-potential field (position-dependent mass, identified also as a Higgs-type field \cite{yann14,roma13}), 
which when integrated into a generalized Dirac equation for the electrons provides a unifying framework for 
the analysis of transport processes through graphene constrictions and segmented junctions. 

With growing activities and further improvements in the areas of bottom-up fabrication and manipulation
of atomically precise \cite{ruff10,huan12,derl13,nari14,sini14} 
graphene nanostructures and the anticipated measurement
of conductance through them, the above findings could serve as impetus and implements 
aiding the design and interpretation of future experiments.\\
~~~~~~~~~~~~\\
{\bf Methods}\\
~~~~~~~~~~~~~~\\
{\small
{\bf Dirac-Fabry-P\'{e}rot model}.
The energy of a relativistic fermion (with one-dimensional momentum $p_x$) is given by the Einstein relation 
$E=\sqrt{ (p_x v_F)^2+({\cal M} v_F^2)^2 }$, where ${\cal M}$ is the rest mass and $v_F$ is the
Fermi velocity of graphene. (In a uniform armchair graphene nanoribbon, the mass parameter is related to the
particle-hole energy gap, $\Delta$, as ${\cal M}=\Delta/(2 v_F^2)$.) Following the relativisitic 
quantum-field Lagrangian formalism, the mass ${\cal M}$ is replaced by a position-dependent scalar 
Higgs field $\phi(x) \equiv m(x)v_F^2$, to which the relativistic fermionic field $\Psi(x)$ couples 
through the Yukawa Lagrangian \cite{roma13} ${\cal L}_Y = -\phi \Psi^\dagger \beta \Psi$ 
($\beta$ being a Pauli matrix). For $\phi(x) \equiv \phi_0$ (constant) ${\cal M} v_F^2=\phi_0$, and 
the massive fermion Dirac theory is recovered. The Dirac equation is generalized as (here we do not
consider applied electric or magnetic fields)
\begin{equation}
[E-V(x)] \Psi + i \hbar v_F \alpha \frac{\partial \Psi}{\partial x} - \beta \phi(x) \Psi=0.
\label{direq}
\end{equation}
In one dimension, the fermion field is a two-component spinor $\Psi = (\psi_u, \psi_l)^T$;
$u$ and $l$ stand, respectively, for the upper and lower component and $\alpha$ and $\beta$ can be any 
two of the three Pauli matrices. Note that the Higgs field enters in the last term of Eq.\ (\ref{direq}).
$V(x)$ in the first term is the usual electrostatic potential, which is inoperative due to the Klein 
phenomenon \cite{kats06,rodr12} and thus is set to zero for the case of the armchair nanoribbons (where the 
excitations are relativistic). The generalized Dirac Eq. (\ref{direq}) is used in the 
construction of the transfer matrices of the Dirac-Fabry-P\'{e}rot model described below.

The building block of the DFP model is a 2$\times$2 wave-function matrix ${\bf \Omega}$ formed by the
components of two independent spinor solutions (at a point $x$) of the onedimensional, first-order
generalized Dirac equation [see Eq.\ (3) in the main paper]. 
${\bf \Omega}$ plays \cite{mcke87} the role of the Wronskian matrix ${\bf W}$ used in the second-order 
nonrelativistic Kronig-Penney model. Following Ref.\ \onlinecite{mcke87} and generalizing to $N$
regions, we use the simple form of ${\bf \Omega}$ in the Dirac representation
($\alpha=\sigma_1$, $\beta=\sigma_3$), namely
\begin{equation}
{\bf \Omega}_K (x) = \left( \begin{array}{cc}
e^{i K x} & e^{-i K x} \\
\Lambda e^{i K x}& -\Lambda e^{-i K x} \end{array} \right),
\label{ome}
\end{equation}
where
\begin{equation}
K^2=\frac{(E-V)^2-m^2 v_F^4}{\hbar^2 v_F^2}, \;\;\; \Lambda= \frac{\hbar v_F K}{E-V+m v_F^2}.
\label{klam}
\end{equation}
The transfer matrix for a given region (extending between two matching points $x_1$ and $x_2$
specifying the potential steps $m_i^{(n)}$) is the product
${\bf M}_K (x_1,x_2)= {\bf \Omega}_K (x_2){\bf \Omega}_K^{-1} (x_1)$;  this latter matrix depends
only on the width $x_2-x_1$ of the region, and not separately on $x_1$ or $x_2$.

The transfer matrix corresponding to a series of $N$ regions can be formed \cite{roma13} 
as the product
\begin{equation}
{\bf t}_{1,N+1} = \prod_{i=1,N} {\bf M}_{K_i} (x_i,x_{i+1}),
\label{tside}
\end{equation}
where $|x_{i+1}-x_i|=L_i$ is the width of the $i$th region [with $(m,V,K,\Lambda) \rightarrow
(m_i,V_i,K_i,\Lambda_i)$]. The transfer matrix associated with the
transmission of a free fermion of mass ${\cal M}$ (incoming from the right) through the 
multiple mass barriers is the product
\begin{equation}
{\bf {\cal T}}(E) = {\bf \Omega}^{-1}_k(x_{N+1}) {\bf t}_{1,N+1} {\bf \Omega}_k(x_1),
\label{thex}
\end{equation}
with $k=\sqrt{(E-V)^2-{\cal M}^2 v_F^4}/(\hbar v_F)$, $|E-V| \geq {\cal M} v_F^2$; for armchair leads
$V=0$, while for zigzag leads $V=\mp {\cal M} v_F^2$. Naturally, in the case of metallic armchair leads, 
$k=E/(\hbar v_F)$, since ${\cal M}=0$.

Then the transmission coefficient $T$ is
\begin{equation}
T=\frac{1}{|{\cal T}_{22}|^2},
\label{trans}
\end{equation}
while the reflection coefficient is given by
\begin{equation}
R=\left| \frac{{\cal T}_{12}}{{\cal T}_{22}}\right| ^2.
\label{refl}
\end{equation}

At zero temperature, the conductance is given by $G = (2e^2/h) T$;
$T$ is the transmission coefficient in Eq.\ (\ref{trans}).\\

{\bf TB-NEGF formalism}.
To describe the properties of graphene nanostructures in the tight-binding approximation, we use the 
hamiltonian 
\begin{equation}
H_{\text{TB}}= - t \sum_{<i,j>} c^\dagger_i c_j + h.c.,
\label{htb}
\end{equation}
with $< >$ indicating summation over the nearest-neighbor sites $i,j$. $t=2.7$ eV
is the hopping parameter of two-dimensional graphene.

To calculate the TB-NEGF transmission coefficients, the Hamiltonian (\ref{htb}) is employed in conjunction 
with the well known transport formalism which is based on the nonequilibrium Green's functions. 
\cite{dattabook}

According to the Landauer theory, the linear conductance is $G(E)=(2 e^2/h) T(E)$,
where the transmission coefficient is calculated as $T(E)=\Tr[\Gamma_L {\cal G} \Gamma_R {\cal G}^\dagger]$.
The Green's function ${\cal G}(E)$ is given by 
\begin{equation}
{\cal G}(E) = (E+i \eta - H_{\text{TB}}^{\text{dev}} - \Sigma_L - \Sigma_R)^{-1},
\label{ttb}
\end{equation}
with $H_{\text{TB}}^{\text{dev}}$ being the Hamiltonian of the isolated device (junction without the
leads). The self-energies $\Sigma_{L(R)}$ are given by 
$\Sigma_{L(R)} = \tau_{L(R)}[E+i \eta - H_{\text{TB}}^{L(R)}]^{-1}\tau_{L(R)}^\dagger$,
where the hopping matrices $\tau_{L(R)}$ describe the left (right) device-to-lead coupling,
and $H_{\text{TB}}^{L(R)}$ is the Hamiltonian of the semi-infinite left (right) lead. 
The broadening matrices are given by $\Gamma_{L(R)}=i[\Sigma_{L(R)}-\Sigma_{L(R)}^\dagger]$.   
} 

~~~~\\
\noindent
{\bf Acknowledgements:} 
{\small This work was supported by a grant from the Office of Basic Energy Sciences of 
the US Department of Energy under Contract No. FG05-86ER45234. Computations were made at the GATECH 
Center for Computational Materials Science.}\\
~~~~~~\\
{\bf Author Contributions:} {\small I.R. \& C.Y. performed the computations. C.Y., I.R. \& U.L. analyzed the 
results. C.Y. \& U.L. wrote the manuscript.}\\
~~~~~~~~~~~~\\
{\bf Competing financial interests:} {\small The authors declare no competing financial interests.}\\
~~~~~~~~~~~~~\\
{\bf Correspondence} {\small should be addressed to C.Y.: (Constantine.Yannouleas@physics.gatech.edu).}

\end{document}